\documentclass[journal=jacsat,manuscript=article]{achemso}

\usepackage[version=3]{mhchem} 
\usepackage{physics}
\usepackage{xcolor}
\usepackage{subcaption}
\usepackage{hyperref}
\usepackage{color}
\usepackage{overpic}
\usepackage{bm}
\usepackage[utf8]{inputenc}

\newcommand{\bg}{\mbox{\protect\boldmath $g$}}

\newcommand{\bu}{\mbox{\protect\boldmath $u$}}

\newcommand{\bp}{\mbox{\protect\boldmath $p$}}

\def\AmS{{\the\textfont2 A}\kern-.1667em\lower.5ex\hbox
     {\the\textfont2 M}\kern-.125em{\the\textfont2 S}}

\def\AW{Addison\kern.1em-\penalty0\hskip0pt Wesley}
\def\BibTeX{{\rm B\kern-.05em{\smc i\kern-.025emb}\kern-.08em\TeX}}
\author{Jakub Martinka}
\affiliation{J. Heyrovsk\'{y} Institute of Physical Chemistry, Academy of Sciences of the Czech \mbox{Republic, v.v.i.}, Dolej\v{s}kova 3, 18223 Prague 8, Czech Republic}
\alsoaffiliation{Department of Physical and Macromolecular Chemistry, Faculty of Science,
Charles University, Hlavova 8, 128 43 Prague 2, Czech Republic}

\author{Mikołaj Martyka}
\affiliation[First University]
{Faculty of Chemistry, University of Warsaw, Pasteura 1, Warsaw, 02-093, Poland}

\author{Biman Medhi}
\affiliation[First University]
{Department of Chemistry, Indian Institute of Technology, Guwahati, Assam, India}

\author{Ji\v{r}\'{i} Pittner}
\affiliation{J. Heyrovsk\'{y} Institute of Physical Chemistry, Academy of Sciences of the Czech \mbox{Republic, v.v.i.}, Dolej\v{s}kova 3, 18223 Prague 8, Czech Republic}

\author{Pavlo O. Dral}
\email{dral@xmu.edu.cn}
\affiliation[Second University]
{State Key Laboratory of Physical Chemistry of Solid Surfaces, Department of Chemistry, College of Chemistry and Chemical Engineering, and Fujian Provincial Key Laboratory of Theoretical and Computational Chemistry}
\alsoaffiliation[Third University]
{Institute of Physics, Faculty of Physics, Astronomy, and Informatics, Nicolaus Copernicus University in Toru\'n, ul. Grudziadzka 5, 87-100 Toru\'n, Poland}
\alsoaffiliation[Aitomistic]
{Aitomistic, Shenzhen 518000, China}

\title[An \textsf{achemso} demo]
  {Flexible Framework for Surface Hopping: From Hybrid Schemes for Machine Learning to Benchmarkable Nonadiabatic Dynamics}
  
\abbreviations{IR,NMR,UV}

\begin{document}

\begin{tocentry}





	\includegraphics[width=\textwidth]{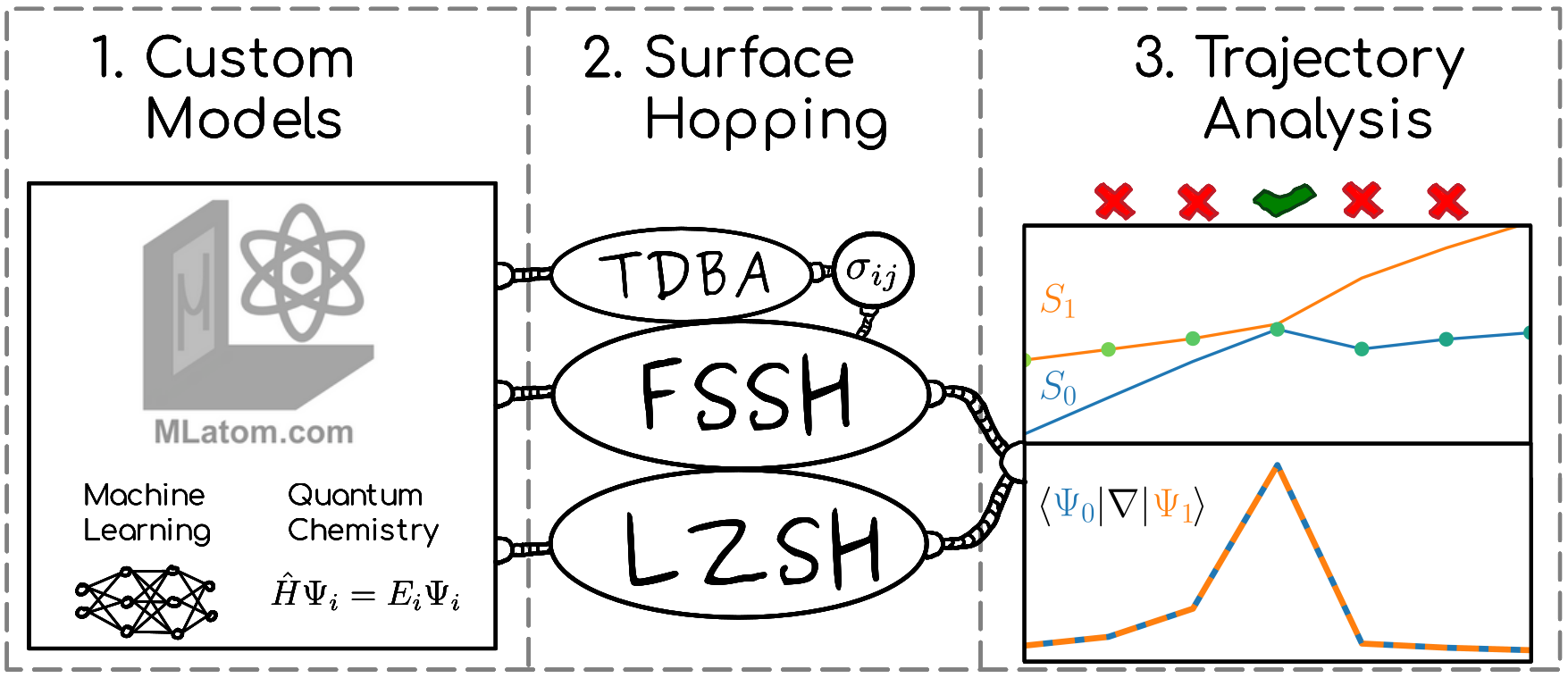}

\end{tocentry}

\begin{abstract}
Nonadiabatic molecular dynamics is a key technique for investigating a broad range of photochemical and photophysical processes. Among the established approaches, surface hopping schemes are widely used and can be easily integrated with various quantum chemistry programs or machine learning models. We present a flexible framework in MLatom that includes a newly implemented Tully's fewest-switches surface hopping algorithm and its time-dependent Baeck--An variant. The capabilities of this framework are demonstrated through three representative examples corresponding to typical stages of a surface hopping study. First, we focus on methods providing energy, energy gradients and nonadiabatic couplings. We show that the flexibility of user-defined custom models can save computational time and that it is useful for benchmarking machine learning models. Next, we compare curvature-driven surface hopping schemes and show that Landau--Zener approach outperforms the time-dependent Baeck--An scheme. Finally, we showcase easy-to-use analysis tools for both individual trajectories and trajectory ensembles. This framework enables accelerated development of machine learning models and provides deeper insight into nonadiabatic dynamics. It is available as a part of the open-source MLatom package.
\end{abstract}

\section{Introduction}
The computational tools are an essential part of chemical research. They allow deep insight into complex processes in atomistic systems, complementing experimental investigation. Unfortunately, every tool, regardless of whether it is experimental or theoretical, is inherently limited to a certain degree. One of the limiting factors is the level of theory that is used within a simulation. In a simplified view of quantum chemistry (QC), the more accurate method or larger basis set we choose, the more precise the result, and more computational time has to be spent. While performing simulations like molecular dynamics, one is therefore forced to sacrifice accuracy to be able to describe a large system or to reduce the time of performing the simulations. Machine learning (ML) helped push these limits by making fast predictions while maintaining the accuracy of the reference training data, and is now a well-established part of computational chemistry\cite{Dral2020}. The MLatom package\cite{MLatom2024} is an open-source computational chemistry toolbox that facilitates this connection between QC methods, state-of-the-art ML models and simulations. It provides a wide range of different simulation techniques, e.g. geometry optimisation, reaction exploration, spectra calculation or molecular dynamics. A variety of interfaces to QC packages and ML models make it a particularly useful tool to study excited-state phenomena. Here, we present our latest development in the direction of nonadiabatic molecular dynamics (NAMD) simulations, particularly various schemes of Tully's fewest-switches surface hopping (FSSH) algorithm, enlarging MLatom's repertoire of simulation techniques.

The FSSH is a popular tool for the investigation of photophysical and photochemical processes in molecular systems, especially due to its simplicity and scalability. It is based on separating the system into classical (nuclei) and quantum (electrons) degrees of freedom. For a given molecular geometry ${\bm R}=\{{\bm R}_A\}_{A=1}^{N_{\mathrm{nucl}}}$, the time-independent electronic Schr\"{o}dinger equation is solved to obtain ground and excited state energies, which serve as a potential for the motion of nuclei obtained by integration of Newton's equations of motion. This leads to a nuclear trajectory on an excited potential energy surface (PES), with the possibility of a stochastic `hop' to a different PES. A large ensemble of these trajectories provides a statistical picture of population transfer between electronic states as well as configurational changes associated with such a process. Since its development in 1990s\cite{Tully1990,HammesSchiffer1994} the large number of software performing FSSH simulations emerged, such as Newton-X\cite{NewtonX2022}, SHARC\cite{Mai2018, SHARC4.0}, PyRAI2MD\cite{Li2021,Li2021a}, PYXAID\cite{Akimov2013}, MNDO\cite{Fabiano2008}, JADE\cite{Du2015}, ANT\cite{YinanShu2025}, Libra\cite{Shakiba2022}, ABIN\cite{Suchan2020}, FISH\cite{Mitric2009}, Hefei-NAMD\cite{Zheng2019}, and others.

The application of ML methods to NAMD is a challenging task. At each classical step, FSSH requires the energies of excited electronic states, their gradients and nonadiabatic couplings (NACs) between them. The PESs are non-differentiable on the conical intersection (CI) seam, and the PESs in its vicinity must be accurately described by the ML model in order to correctly reproduce the excited-state deactivation process. Therefore, special care must be taken when assembling the training set, e.g. by employing active learning (AL) strategies\cite{Westermayr2019,Li2021,Westermayr2022,Martyka2025}. Energy gradients are typically obtained from the energy model as analytical derivatives, and they are commonly included in the training procedure. Last quantity to fit, and probably the most problematic one, is the first-order nonadiabatic coupling between states $i$ and $j$ defined as 
\begin{equation}\label{eq:NACs}
{\bm d}_{ij} = \bra{\Psi_i}\nabla_{\bm R}\ket{\Psi_j} = \frac{{\bm h}_{ij}}{E_i-E_j} \quad \forall\, i\neq j,
\end{equation}
where ${\bm h}_{ij}=\bra{\Psi_i}\nabla_{\bm R}\hat{H}_{\text{el}}\ket{\Psi_j}$ and $\Psi_i$ represents an electronic state with corresponding energy $E_i$ for geometry $\bm R$.
The first difference is that energies are scalars, so they are rotationally invariant quantities, while NACs are of a vectorial nature. This means that NACs have to transform equivariantly with respect to the transformation of the input structure, posing a constraint for ML models. Next, when $E_i-E_j\rightarrow0$, see Eq.~\ref{eq:NACs}, the NACs diverge to infinity. Last, due to the arbitrary phase of wavefunctions $\Psi_i$, the sign of NAC corresponding to one geometry $\bm R$ is not well defined. This means that having a data set is not enough for ML to be successful; one has to phase-correct the data or train the models in a way to remove dependence on sign. 

The complicated nature of NACs and their unavailability for some QC methods led to the development of a numerical approach based on the overlap integral of electronic wave functions\cite{HammesSchiffer1994}. This approach has been used in FSSH simulations based on CC2, ADC(2), TDDFT, as well as semiempirical methods\cite{Granucci2001,Tapavicza2007,Mitric2008,Barbatti2010a,Plasser2014,Mai2017}. However, it  cannot be combined with ML models, and several schemes based on the topography of PESs, the so-called curvature-driven surface hopping schemes, have been developed. These schemes require only energies and energy gradients, reducing the problem to fitting the PESs. Nevertheless, fitting of excited PESs is still challenging due to the existence of the CIs, but it builds on the years of experience with fitting ground-state PESs, and high-quality extensions have been successfully reported\cite{Barrett2025,Mausenberger2024a,Martyka2025}.

The MLatom targets an all-in-one strategy that unifies robust ML protocols, NAMD simulation techniques, and convenient analysis of results in a single, versatile Python API. It provides a simple, flexible and easy-to-use framework to study nonadiabatic processes. The MLatom's flexibility lies in the freedom of defining user-specific models combining different QC and ML methods, speeding up both method development and simulations in the production phase. This might be useful for the development of robust ML frameworks and for benchmarking of different surface hopping schemes. We further provide a variety of applications and assessments. Our newly implemented surface hopping schemes, as described in the sections below, are available in MLatom~3.

\begin{figure}
    \centering
    \includegraphics[width=.9\textwidth]{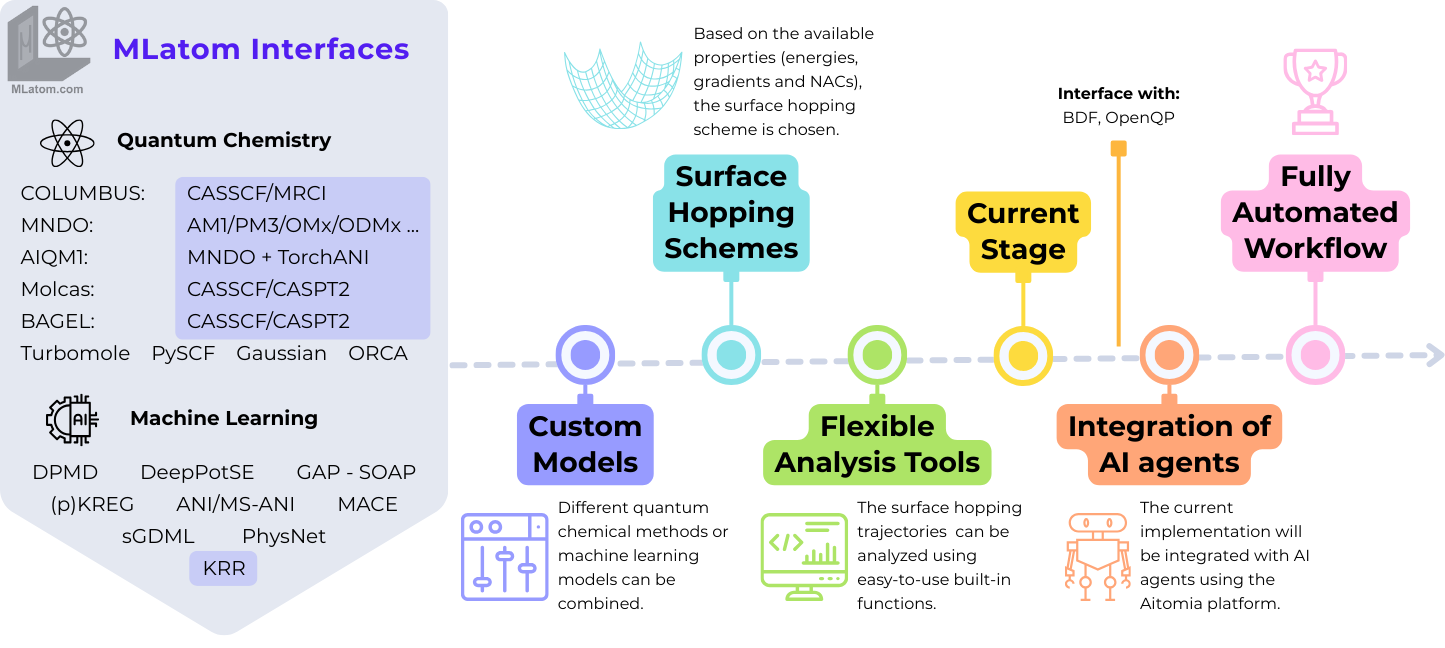}
    \caption{
    The roadmap of NAMD development in MLatom. The highlighted methods in the interfaces section provide NACs and can be used in the FSSH scheme. The custom models, surface-hopping features, and analysis tools are presented in this work. The next step is the development of automated protocols integrated with AI agents.}
    \label{fig:roadmap}
\end{figure}

The current progress and future plans for NAMD implementations in MLatom are shown in the roadmap (Fig.~\ref{fig:roadmap}). At the time of writing, MLatom supports interfaces to the growing number of the QC programs including COLUMBUS\cite{Columbus7.2,Lischka2020}, MNDO\cite{MNDO}, MOLCAS\cite{Veryazov2004,Aquilante2015,LiManni2023}, BAGEL\cite{Bagel,Lehtola2022}, Turbomole\cite{TURBOMOLE_7.5}, PySCF\cite{Sun2017,Sun2020}, Gaussian\cite{Gaussian16} and ORCA\cite{Neese2011}).
These interfaces enable simulations with various methods, such as AIQM1\cite{Zheng2021}; complete active space self-consistent field (CASSCF) and second-order complete active space perturbation theory (CASPT2), in their state-averaged (SA), multi-state (MS), or extended multi-state (XMS) variants; multireference configuration interaction (MRCI); and semiempirical approaches (AM1\cite{Dewar1985}, PM3\cite{Stewart1989}, OMx\cite{Dral2019}, ODMx\cite{Dral2019}, etc.). Furthermore, single-reference second-order algebraic diagrammatic construction (ADC(2)) and time-dependent density functional theory (TDDFT) are also available.
In addition, MLatom's broad library of supported ML models provides a versatile environment for their integration into the NAMD applications; these models include DPMD\cite{Zhang2018a}, DeepPotSE\cite{Zhang2018}, GAP\cite{Bartok2010}--SOAP\cite{Bartok2013}, KRR\cite{Martinka2024}, (p)KREG\cite{Hou2023}, ANI\cite{Smith2017,Gao2020}, MS-ANI\cite{Martyka2025}, MACE\cite{Batatia2022}, sGDML\cite{Chmiela2018}, and PhysNet\cite{PhysNet2019}. Other interfaces such as to the OpenQP\cite{Mironov2024} and BDF\cite{Liu2003,Liu1997,Liu2004,Zhang2020b} QC programs are envisioned in the near future. We also plan to integrate above functionality for autonomous FSSH simulations with AI agents through Aitomia\cite{Hu2025}.

\section{Methods}
\subsection{Fewest-Switches Surface Hopping Dynamics}
There is a plethora of semiclassical methods that can approximate quantum dynamical treatment of systems of nuclei and electrons undergoing nonadiabatic processes, when two or more electronic excited states are involved and the Born--Oppenheimer approximation breaks down.
We focus on the FSSH method, in which the system in the active state $l$ is propagated on PES by solving classical Newton's equations of motion
\begin{equation}
m_A\frac{\mathrm{d}^2{\bm R}_A}{\mathrm{d}t^2}={\bm F}_A=-\nabla_{{\bm R}_A} E_l,
\end{equation}
where force ${\bm F}_A$ acting on nuclei $A$ is the negative gradient of electronic energy $E_l$ obtained by solving time-independent Schrödinger equation for a particular configuration ${\bm R}$.
The coefficients $\{c_k\}_{k=1}^{N}$ of all $N$ adiabatic states are propagated between classical time steps $\Delta t$ using a smaller quantum time step $\Delta\tau$ according to locally approximated \mbox{time-dependent} Schrödinger equation
\begin{equation}\label{eq:scse}
\dot{c}_k = -\sum_{j\neq k} c_j e^{i\gamma_{kj}} \dot{{\bm R}} \cdot {\bm d}_{kj},
\end{equation}
where $\gamma_{kj}$ is a difference between phases $\gamma_k$ and $\gamma_j$ defined as
\begin{equation}
\gamma_{i}=\int_0^tE_i({\bm R}(t'))\mathrm{d}t'.
\end{equation}

For each quantum time step, a random number $r$ is uniformly generated from an interval of $[0,1]$ and a so-called `hop' can occur from the active state $l$ to a state $k$ if the condition
\begin{equation}
\sum_{j=1}^{k-1}P_{l\rightarrow j}(t) < r \le \sum_{j=1}^{k}P_{l\rightarrow j}(t)
\end{equation}
is satisfied and if there is enough kinetic energy for total energy conservation in case of $l<k$.
The probability $P_{l\rightarrow j}$ is calculated as
\begin{equation}\label{eq:prob}
P_{l\rightarrow j}^{\text{FSSH}}(t) = \text{max}\left[0,\frac{-2\Delta\tau\text{Re}(a^*_{jl}e^{i\gamma_{jl}(t)})\sigma_{jl}}{a_{ll}(t)}\right],
\end{equation}
where $a_{ij}=c_ic^*_j$ and $\sigma_{jl} = \dot{{\bm R}}\cdot {\bm d}_{jl}$ is the time derivative coupling (TDC).

Various corrections to the state coefficients $c_k$ were proposed to account for the decoherence, e.g., based on overlap\cite{Granucci2010} or decay of mixing\cite{Zhu2004, Hack2001}. In MLatom, we implemented the FSSH with simplified decay of mixing (SDM)\cite{Granucci2007}, where coefficients are transformed as
\begin{subequations}
\begin{equation}
c_k^{'}=c_k\exp\left( \frac{-\Delta\tau|E_{k}-E_{l}|}{1+\alpha/E_{\text{kin}}} \right) \quad \forall\, k\neq l,
\end{equation}
\begin{equation}
c_l^{'}=c_l\sqrt{\frac{1-\sum_{k\neq l}|c_k^{'}|}{|c_l|^2}},
\end{equation}
\end{subequations}
where $l$ is a label of active state and $\alpha$ is a constant with a default value of 0.1~Hartree. The FSSH employing decoherence correction is therefore sometimes called DC-FSSH, but since we always use this correction in our examples, the prefix is omitted.

Surface hopping simulations must conserve the total energy $\Delta E^{\mathrm{tot}}=0$\cite{Tully1991,Herman1995,Herman1984}. This implies that hop from state $l$ to $j$, which is associated with a change in potential energy $\Delta E_{lj}$, must be compensated by change in kinetic energy, so that $\Delta K_{lj}=-\Delta E_{lj}$. This is facilitated by rescaling velocities $\dot{{\bm R}}$ in a general direction $\bm u_A$ according to
\begin{equation}\label{eq:velrescale}
    \dot{{\bm R}}_A^j = \dot{{\bm R}}_A^l + \kappa_{lj}\frac{\bm u_A}{M_A},
\end{equation}
where $\kappa_{lj}$ is a rescaling factor. It is obtained as the lower root of a quadratic equation as described in Supporting Information of Ref.~\citenum{Barbatti2021}. When hopping into an energetically higher state, $l<k$, there has to be enough kinetic energy available; otherwise, the hopping is blocked and called ``frustrated''. The choice of $\bu_A$ depends on the availability of ${\bm d}_{jl}$. The momentum $\bp$ or the gradient difference ${\bg}_{jl}$ are alternatively used, but other schemes based on atomic contribution to the electronic transition were proposed\cite{SangiogoGil2025}.

\subsection{Schemes omitting evaluation of nonadiabatic couplings}
Calculation of the hopping probability (Eq.~\ref{eq:prob}) requires TDCs, which are typically obtained as a scalar product of velocities and NACs. Nevertheless, NACs are not available for some QC methods, like CC2 or ADC(2), and consequently for ML models trained only on energies and gradients. Even when the NACs are included in the training data, their learning represents a non-trivial task. There is thus a strong demand for NAC-free surface hopping schemes, i.e. Zhu--Nakamura\cite{Zhu1994,Zhu1995,Ishida2017}, Landau–Zener surface hopping\cite{Zener1930, Landau1932, Landau1932a,Belyaev2011,Belyaev2014} (LZSH), and time-dependent Baeck–An\cite{Baeck2017,Casal2022} (TDBA, or equivalent scheme called curvature-driven trajectory surface hopping\cite{Shu2022}). 

The TDBA scheme employs the Eq.~\ref{eq:prob} for the probability of hopping as in FSSH, but the TDC is approximated as
\begin{equation}\label{eq:tdba}
    \sigma_{jl} = \frac{1}{2}\sqrt{\frac{\frac{\mathrm{d}^2}{\mathrm{d}t^2}\Delta E_{jl}}{\Delta E_{jl}}}  \quad \forall\,  l > j,
\end{equation}
where $\Delta E_{jl}$ is an potential energy difference between adiabatic states $j$ and $l$.
The TDBA couplings are implemented according to Ref.~\citenum{Casal2022} using the second-order finite differences of potential energies with the backward $O(\Delta t)$ and $O(\Delta t^2)$ approximation. The $\sigma_{jl}$ is set to $-\sigma_{lj}$ if $l < j$ and to zero in case of a negative argument of the square root.

The LZSH approach offers a simpler alternative, as it does not require propagating the state coefficients using Eq.~\ref{eq:scse}; instead, the hopping probability is directly calculated as
\begin{equation}\label{eq:lzbl_prob}
P_{l\rightarrow j}^{\text{LZSH}} = \exp\left(-\frac{\pi}{2\hbar}\sqrt{\frac{\Delta E_{lj}^3}{\frac{\mathrm{d}^2}{\mathrm{d}t^2}\Delta E_{lj}}}\right).
\end{equation}
The denominator in the square root of Eq.~\ref{eq:lzbl_prob} is typically computed from a three-point finite-difference formula for the middle point if it represents a minimum. This means that for the evaluation of the probability at time $t$, one needs to access information at time $t+dt$. 

The curvature-based hopping schemes may suffer from discontinuities in PESs, causing erroneous hopping, which can be mitigated by introducing hop-blocking algorithms\cite{Jira2025}, in the case of LZSH, or conditions imposed on values of TDBA couplings\cite{Casal2022}. In our implementation, we omit those schemes as they tend to be system-specific, and the main interest of MLatom is the application of ML models that provide smooth PESs. Both the TDBA and LZSH schemes have already been combined with ML models to study photochemical problems\cite{Wang2024,Li2025,Martyka2025,Bispo2025}.

\subsection{Trajectory analysis in MLatom}
From the typical surface hopping trajectory, the time evolution of nuclear and electronic degrees of freedom is obtained. A single trajectory is typically of limited significance; nevertheless, it can provide chemical insight into the nonadiabatic process and help identify nonphysical behaviour when designing ML models, such as unexpected dissociation or erroneous hopping. Plotting certain quantities along a single trajectory, e.g. potential energy, adiabatic population or TDCs/NACs, reveals ML model's effects on the trajectory and uncovers its limitations. Moreover, such an analysis can be useful for benchmarking different surface hopping schemes, e.g. when total energy is not conserved, the resulting discontinuity can induce hops across large energy gaps in curvature-driven schemes.

More informative is the analysis of an ensemble of trajectories, e.g., the time evolution of specific degrees of freedom can uncover the reaction channel or relaxation mechanism. It is possible to check the internal consistency of FSSH by comparing the occupation and the average adiabatic population\cite{Granucci2007}. The occupation, defined as a ratio of trajectories in state $i$ ($N_i$) and the total number of trajectories ($N_\mathrm{traj}$), should be close to the adiabatic population of the $i$-th state ($|c_i|^2$) averaged over all trajectories, i.e.
\begin{equation}\label{eq:population}
    \frac{N_i(t)}{N_\mathrm{traj}}=\frac{1}{N_\mathrm{traj}}\sum^{N_\mathrm{traj}}_{j=1}|c^j_i(t)|^2 \quad\forall\, i,t.
\end{equation}
In a typical FSSH study, only the occupation is plotted, and we therefore refer to occupation simply as population unless explicitly stated. The number of trajectories in the ensemble determines the uncertainty. From the population of state $i$ at time $t$ the error margin $\epsilon$ of the 95\% confidence interval can be obtained as
\begin{equation}\label{eq:confidence}
    \epsilon(t) = 1.96\sqrt{\frac{p_i(t)(1-p_i(t))}{N_\mathrm{traj}}}.
\end{equation}
The characteristic time scale $\tau$ of a nonadiabatic process can be extracted by fitting the population using the proposed kinetic model. In our case of the ferro-wire example, we simply fit an exponential function
\begin{equation}
f(t) = A(1 - e^{-t/\tau}),
\end{equation}
where $A$ is related to the infinite-time population limit.

\begin{figure}
    \centering
    \includegraphics[width=.5\textwidth]{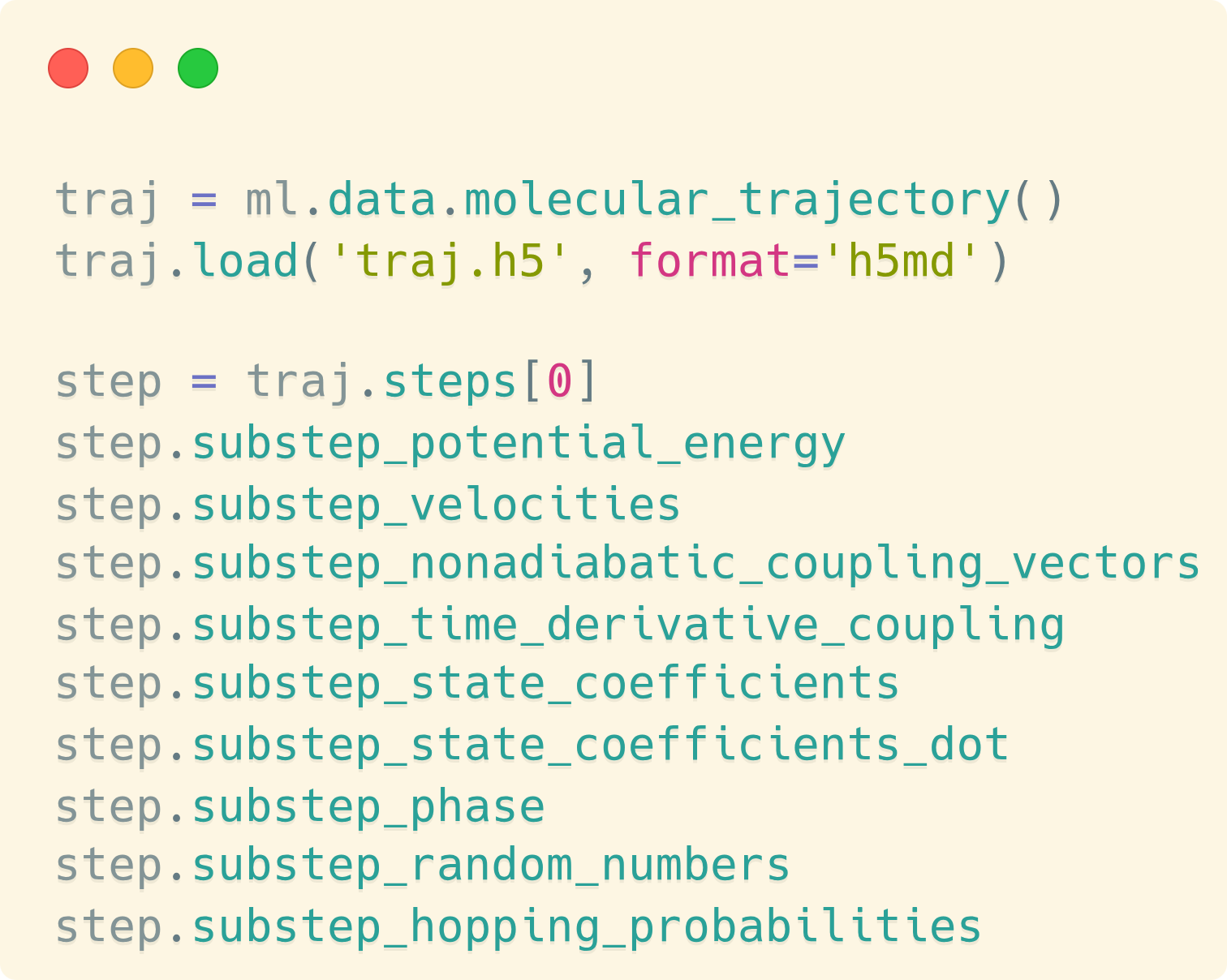}
    \caption{Code snippet example of accessing trajectory information. The \texttt{molecular\_trajectory} instance contains a list of classical time steps. Each \texttt{step} instance contains all necessary data, such as interpolated potential energies, velocities, NACs, TDCs, state coefficients and their derivatives, phases, hopping probabilities and random numbers.}
    \label{fig:code5}
\end{figure}

The principal concern for the analysis of surface hopping trajectories in MLatom is the corresponding data structure. To access the data within MLatom, one has to load the \texttt{molecular\_trajectory} instance containing information about every classical time step. In the case of FSSH or TDBA, every \texttt{step} instance contains the data of the consecutive quantum steps (see Fig.~\ref{fig:code5}), namely, interpolated potential energies, velocities, NACs as well as calculated TDCs, state coefficients and their derivatives, phases, hopping probabilities and corresponding random numbers. The trajectories can be saved into and loaded from a memory-efficient binary H5MD file\cite{Buyl2014}.

When analysing trajectories, the surface hopping scheme determines the quantities available for plotting. In the case of FSSH, both the magnitude of NACs and TDCs can be plotted. In the case of TDBA, only the TDCs are available, while there is no quantum step data for LZSH. The different outputs (on Fig.~\ref{fig:analysis}) are therefore produced using \texttt{plot\_trajs()} method, when such an analysis is performed. When NACs are available, as in the case of FSSH, it might be useful to visualise the Frobenius norm defined as
\begin{equation}
    ||{\bm A}||_F = \sqrt{\sum_{i=1}^m\sum_{j=1}^n|{\bm A}_{ij}|^2}.
\end{equation}

\section{Applications}
In this section, we demonstrate MLatom’s flexibility by presenting three examples that highlight different aspects of surface hopping simulations. First, we address the initial step of any NAMD study: selecting a method capable of providing the required electronic quantities. This is not a trivial task, as numerous QC software packages exist, each offering various methods at different levels of theory. One must therefore strike a delicate balance between accuracy and computational cost. Here, we focus on the advantages of using custom models, which allow users to combine different QC methods or ML models to either accelerate trajectory propagation or investigate ML model's performance. 
Second, once the method providing all necessary quantities is chosen, the next step is to select the appropriate NAMD scheme based on the availability of NACs or TDCs. MLatom offers several options: FSSH, LZSH, and TDBA. When ML models or QC methods do not supply NACs, one must choose between the LZSH and TDBA schemes. We compare these schemes for QC methods as well as the ML models.
Finally, we turn to the analysis of the resulting trajectories. This is a crucial stage, particularly when using ML models that may yield inaccurate results in the extrapolation regime. It is therefore essential to provide user-friendly tools for trajectory and ensemble analysis. MLatom leverages Python’s plotting capabilities to easily visualise properties along trajectories forming a cohesive ecosystem that streamlines benchmarking of ML-accelerated surface hopping. This is demonstrated on examples of fulvene, molecular ferro-wire and methylenimmonium cation.

\subsection{Fulvene: Use of custom models}
\begin{figure}
    \centering
    \includegraphics[width=.9\textwidth]{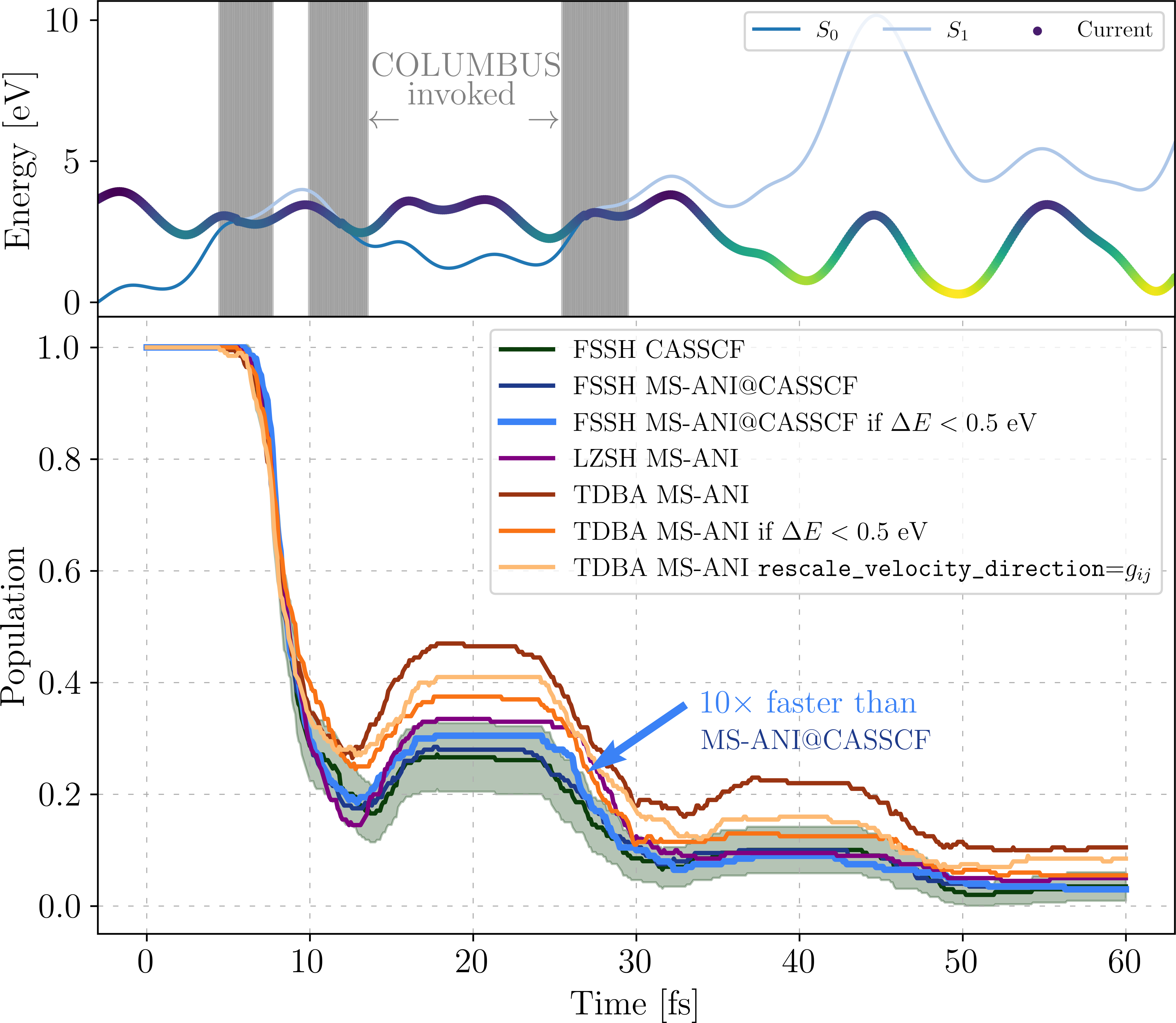}
    \caption{
    Population comparison between different surface hopping schemes for 200 trajectories of the fulvene using custom models. The 95\% confidence interval is shown for FSSH at SA-2-CASSCF(6,6)/6-31G* in Columbus (dark green). The MS-ANI model predicting energies and gradients, combined with CASSCF NACs (dark and light blue), referred to as PES@MS-ANI+NACs@CASSCF, agrees with the CASSCF reference, demonstrating the transferability of AL from LZSH to FSSH. Custom models can be easily modified to incorporate user-defined conditions, such as calculating NACs/TDCs only in a nonadiabatic region (light blue and orange). This approach can significantly speed up the simulations and provides an opportunity for benchmarking surface hopping schemes and ML models. The upper panel shows a typical FSSH PES@MS-ANI+NACs@CASSCF trajectory, where Columbus is called to calculate NACs only when the $S_1$ and $S_0$ states are closer than 0.5 eV (vertical grey lines). Comparing LZSH and TDBA to FSSH shows that LZSH (purple) lies within the confidence interval most of the time and outperforms TDBA (dark red), even when velocity rescaling is done along gradient difference $g_{ij}$ (soft orange) or when the condition to compute TDCs only in nonadiabatic region is applied (orange).}
    \label{fig:pop_f}
\end{figure}

The first example is fulvene, which undergoes isomerisation of the C=C double-bond after excitation into the first excited singlet state. It was suggested as a molecular representative of Tully model III\cite{Ibele2020,Gomez2024} for its reflection process, see Fig.~\ref{fig:pop_f}, due to the nearly degenerate PESs, and was used for benchmarking velocity rescaling schemes\cite{Barbatti2021,SangiogoGil2025}. Even from the perspective of NAMD algorithms, it is a well-studied system, starting from works using variational multi-configurational Gaussian wave packet\cite{MendiveTapia2010,MendiveTapia2012}, through ab-initio multiple spawning\cite{Ibele2020,Mannouch2024}, mapping approach to surface hopping\cite{Mannouch2024}, up to FSSH or LZSH employing ML models\cite{Martinka2025,Martyka2025} as well as TDBA\cite{Casal2022,Bispo2025}. 

\begin{figure}
    \centering
    \includegraphics[width=.95\textwidth]{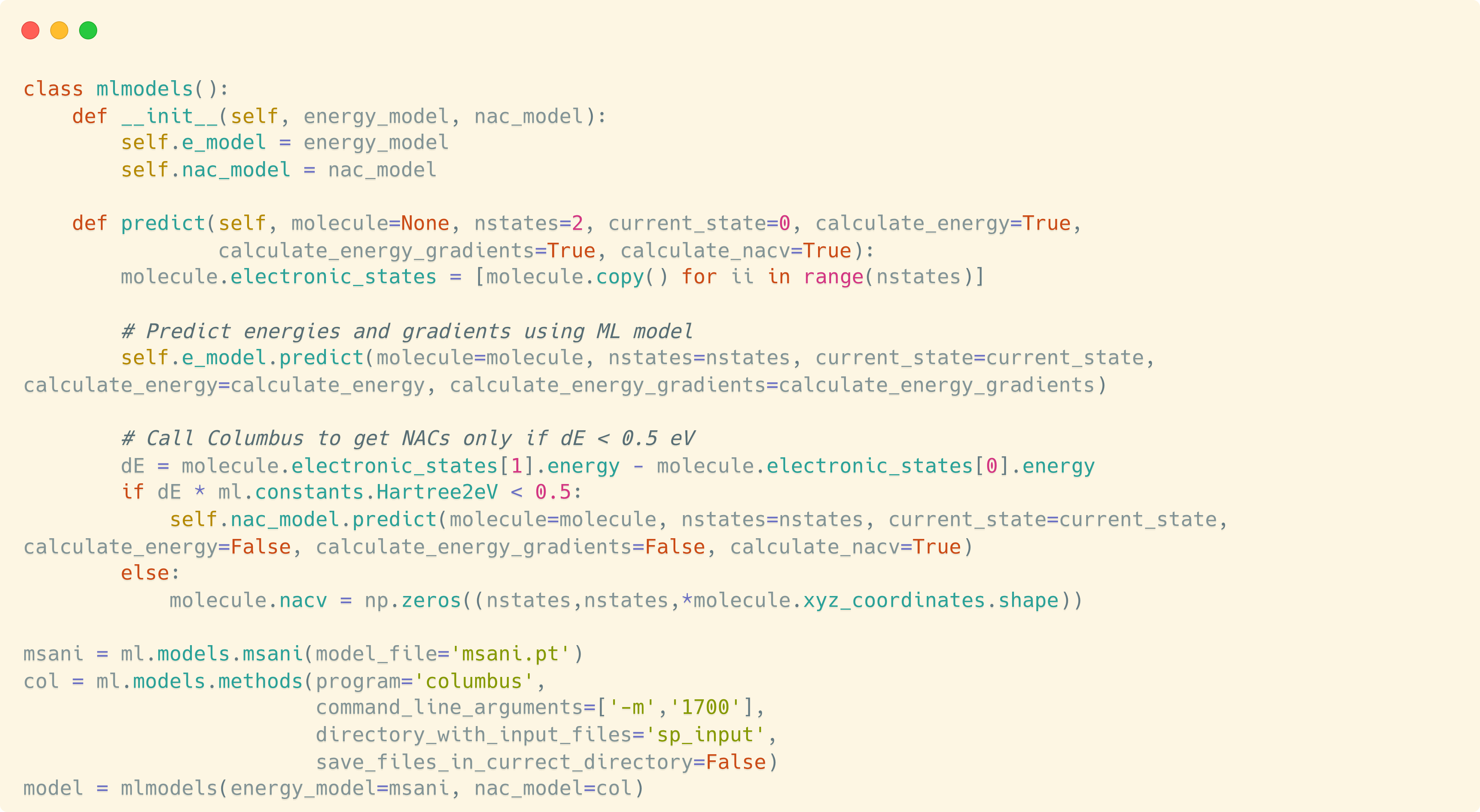}
    \caption{A code snippet example of defining a custom class in MLatom, which combines QC method (CASSCF in Columbus) and ML model (MS-ANI model). The condition can be defined to use CASSCF to calculate NACs only in nonadiabatic region, e.g. when $\Delta E<0.5$ eV.}
    \label{fig:code4}
\end{figure}

To run the FSSH trajectory, it is required to provide energies, gradients and NACs. MLatom has many interfaces to third-party programs as well as pretrained or trainable ML models, making it easy to combine them. It is possible to combine two ML models together, for example, multi-state ANI (MS-ANI~\cite{Martyka2025}) for energies and gradients and kernel ridge regression for NACs. Nevertheless, learning of NACs requires taking care of vector-covariances, e.g. by the rotate-predict-rotate approach~\cite{Martinka2024} or equivariance~\cite{Grisafi2018,Mausenberger2024a}, and the use of NAC-specific descriptors\cite{Martinka2025}. In this example, we show the combination of an MS-ANI model taken from Ref.~\citenum{Martyka2025} for energies and gradients, while NACs are loaded from a SA-2-CASSCF(6,6)/6-31G* calculation with Columbus (version 7.2, 2022)\cite{Lischka2020,Columbus7.2}. The motivation for such a combination is not to speed up the calculation, but rather to demonstrate the convenience of combining different methods and the associated benchmarking potential. The combination is realised by defining a custom class with a \texttt{predict()} method, which in our case requires only 8 lines of Python code. The script running surface hopping further contains loading initial conditions, \texttt{energy\_model} and \texttt{nac\_model}, defining surface hopping settings (\texttt{time\_step}, \texttt{maximum\_propagation\_time}, \texttt{hopping\_algorithm}, \texttt{initial\_state} or \texttt{rescale\_velocity\_direction}), and running and dumping trajectories (see Fig.~\ref{fig:code}). There is no need to recompile the code, and one is free to experiment with different settings, for example, by defining a condition that NACs will be used only when the difference between PESs is less than 0.5 eV\cite{Cui2014,Akimov2018}, see demonstration of such a class in Fig.~\ref{fig:code4} and complete code in repository corresponding to this work: \url{https://github.com/JakubMartinka/FSSH-in-MLatom}. This condition can speed up simulations significantly; it is even possible to combine different levels of theory for PESs/NACs as described in other works, e.g., in the case of TDDFT/TDA\cite{WasifBaig2021} or in the context of ML-accelerated FSSH~\cite{Westermayr2022} and ground-state dynamics with MLatom~\cite{Zhang2023}.

From the resulting trajectories, the population can be obtained by calling the \texttt{plot\_population()} method on a list of \texttt{molecular\_trajectory} instances. In Fig.~\ref{fig:pop_f}, we show the first excited-state decay obtained with different surface hopping schemes using the MS-ANI model and/or the CASSCF method. The MS-ANI model results from MLatom's AL scheme, which efficiently samples small-gap regions using an uncertainty measure and gap-driven dynamics. It is important to point out that this sampling was performed employing only the LZSH scheme, so no information about the FSSH probability based on NACs is provided; the training details can be found in Ref.~\citenum{Martyka2025}. We used the custom model to combine MS-ANI, which predicts energies and gradients, with CASSCF, which supplies NACs. This PES@MS-ANI+NACs@CASSCF combination (dark blue line) yields excellent agreement with the CASSCF reference (dark green line), indicating that the AL scheme can produce a training set that is transferable across different surface hopping schemes. In other words, the MS-ANI model trained on the resulting training set produces accurate PESs, and FSSH can be performed once reliable NACs are provided. Therefore, fitting NACs for FSSH can be treated as a separate task. A significant speed-up is achieved by calculating NACs only when PESs are close to each other. When this condition is applied, the population remains within a statistical error of the CASSCF reference, while only one-tenth of the original time is required to complete all trajectories, making it a potentially useful and easy-to-use exploratory approach. 

\subsection{Molecular ferro-wire: Benchmarking different surface hopping schemes}
As a second example, we compare different curvature-driven surface hopping schemes, namely LZSH and TDBA, against FSSH for a molecular ferro-wire, a large system consisting of 80 atoms with four excited singlet states. This system has previously been studied with FSSH at OM2/MR-CISD\cite{Jankowska2019}, ODM2/CIS, and ODM2/MR-CISD\cite{Zhang2024} level of theory in MNDO\cite{Fabiano2008}, as well as with LZSH using semiempirical methods, AIQM1\cite{Zheng2021}, and ML models\cite{Zhang2024,Martyka2025}.

\begin{figure}
    \centering
    \includegraphics[width=.9\textwidth]{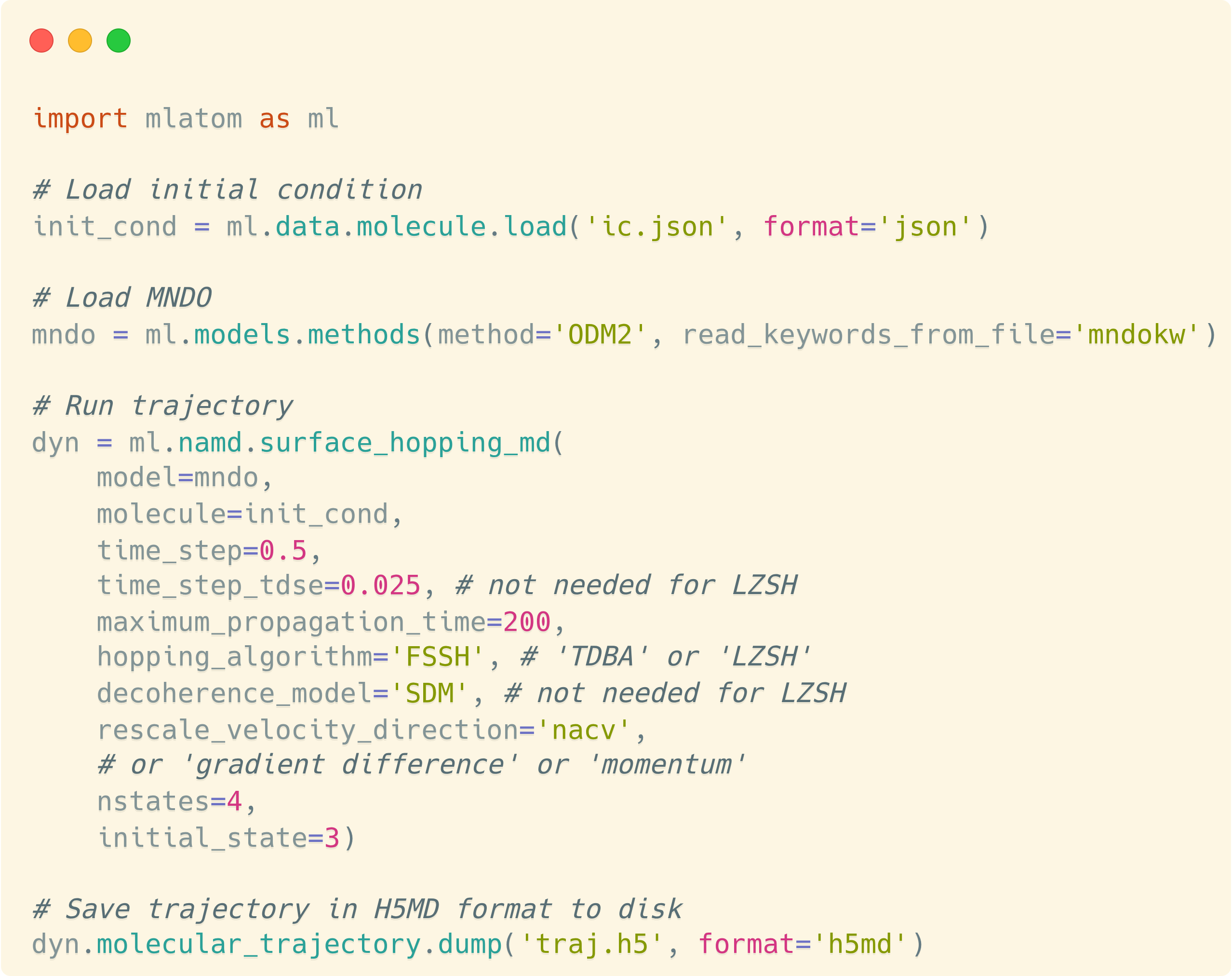}
    \caption{
    Code snippet example of a molecular ferro-wire surface hopping trajectory simulation in MLatom. The initial condition is loaded from a JSON file, and the method providing the necessary information (OM2/MR-CISD in MNDO) is defined. The \texttt{hopping\_algorithm} can be set to FSSH, TDBA, or LZSH, along with other simulation parameters such as time step, time step of state coefficients propagation, maximum propagation time, decoherence model, initial state or velocity rescaling direction.
    }
    \label{fig:code}
\end{figure}

As shown in Fig.~\ref{fig:code}, the surface hopping scheme is specified using the \texttt{hopping\_algorithm} keyword and can be easily changed together with the velocity-rescaling direction. This makes the MLatom's NAMD module a potentially useful tool for benchmarking different surface hopping schemes, as it is convenient to perform them under identical settings and initial conditions. Furthermore, once accurate ML models have been trained, the simulations become extremely fast, enabling large ensembles of trajectories to be run under different parameters to investigate their impact on the results (as demonstrated in the case of fulvene). The data are stored in a unified format, allowing the resulting trajectories to be compared with standardised analysis tools (see next section). A total of 272 trajectories were initiated in the excited state, according to oscillator strengths as described in Ref.~\citenum{Jankowska2019}, with a time step of 0.5 fs and a final time of 200 fs. All trajectories were generated with the MLatom surface hopping implementation interfaced with MNDO using ODM2/MR-CISD level of theory. A small number of trajectories exhibited violations of total energy conservation and were excluded from the analysis.

\begin{figure}
    \centering
    \includegraphics[width=\textwidth]{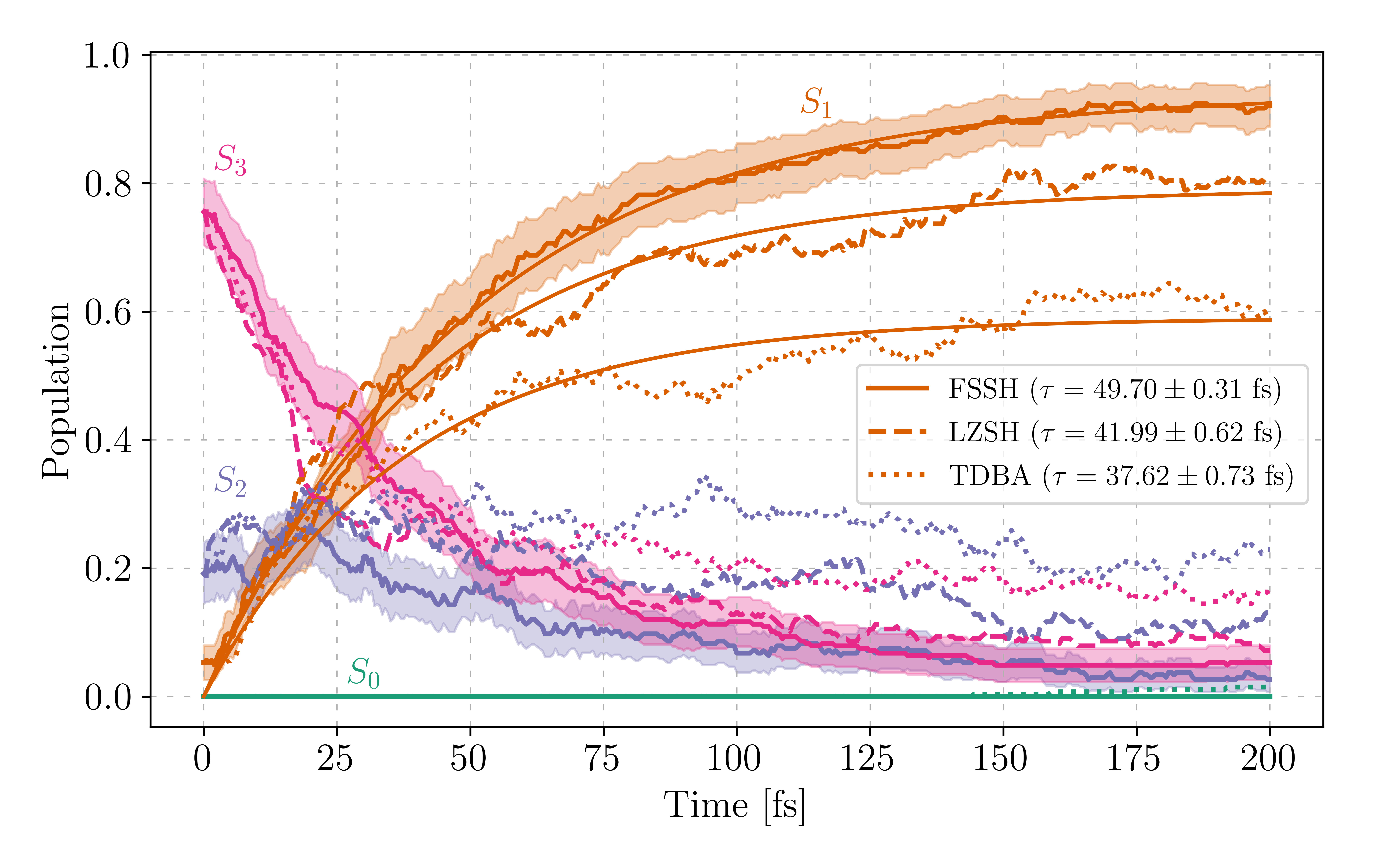}
    \caption{Population comparison of the molecular ferro-wire between LZSH and TDBA, against FSSH at ODM2/MR-CISD level of theory. Totally 272 trajectories were initiated, while the total-energy nonconserving trajectories were removed from the ensemble, resulting in 266 (FSSH), 267 (LZSH), 270 (TDBA) trajectories. The LZSH outperforms TDBA. The $S_1$ population converges to a different final value, and the resulting time scale $\tau$ also differs, with LZSH remaining closer to the FSSH reference.
    }
    \label{fig:fw_pop}
\end{figure}

From Fig.~\ref{fig:fw_pop}, which shows the population of ferro-wire (and from Fig.~\ref{fig:pop_f} and Fig.~\ref{fig:analysis}a, which show fulvene and methylenimmonium cation, respectively), we confirm that LZSH outperforms TDBA in all cases. The $S_1$ population converges to a different final value, and the resulting time scale $\tau$ also differs, with LZSH remaining closer to the FSSH reference. LZSH and TDBA are based on the PES's curvature and suffer from discontinuities. Looking at the individual trajectories, it is not easy to detect the effect of such discontinuities and their influence on hopping probability, particularly given the stochastic nature of surface hopping. In the case of LZSH, the analysis is more straightforward due to the hopping probability being calculated directly from the energy gap and its second derivative (Eq.~\ref{eq:lzbl_prob}). On the other hand, in the case of TDBA, the effect of PES discontinuities transfers to TDCs, which are only one component of the hopping-probability formula (Eq.~\ref{eq:tdba}). The analysis of trajectories did not reveal any evident effects of such discontinuities.

\begin{figure}
    \centering
    \includegraphics[width=\textwidth]{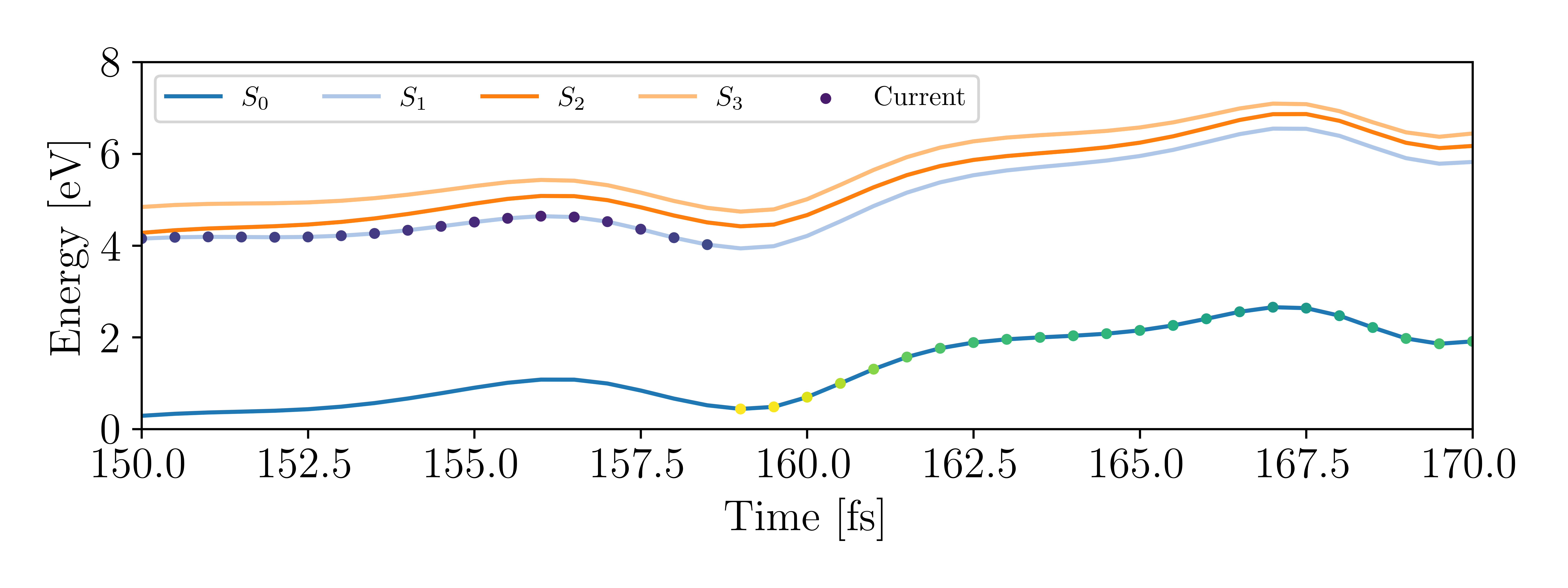}
    \caption{
    Molecular ferro-wire surface hopping trajectory using TDBA couplings showing erroneous hopping from $S_1$ to $S_0$ over an energy gap of 3.5 eV. The condition to calculate TDBA couplings only when $\Delta E_{ij}<0.5$ eV would mitigate such hopping.}
    \label{fig:tdba_hop}
\end{figure}

Another argument in favour of LZSH over TDBA is that Fig.~\ref{fig:fw_pop} shows a non-zero population of the $S_0$ state in the case of the TDBA scheme. This indicates that unphysical hopping occurred in some trajectories, as the energy gap between $S_0$ and $S_1$ ranges from 2.8 to 4.5 eV. Indeed, this happened in three trajectories, where the average energy gap at the hopping point was 3.7 eV (see Fig.~\ref{fig:tdba_hop}). Such unphysical hopping would not occur if the condition to calculate TDCs only when $\Delta E_{ij}<0.5$ eV was employed. Neither the energy-gap condition nor constraints imposed on TDC values would likely increase the population transfer to $S_1$. Curvature-driven surface hopping schemes were recently benchmarked for model Hamiltonians and realistic systems\cite{Jira2025}, suggesting that the LZSH scheme may be preferred over TDBA. The authors point out that one of the main limitations of TDBA is the underestimation of TDC peaks, which explains why TDBA yields slower $S_2\rightarrow S_1$ population transfer in the case of the ferro-wire system. This aligns with another study reporting incorrect TDBA dynamics in certain cases\cite{Merritt2023}. The examples presented in this work (fulvene, molecular ferro-wire, and the methylenimmonium cation) support the conclusion that LZSH should be preferred over TDBA; however, we did not implement any additional conditions to control the values of TDCs\cite{Casal2022}, nor a blocking algorithm for discontinuity-induced hops\cite{Jira2025}.

\subsection{Methylenimmonium cation: Trajectory analysis}
Regardless of the surface hopping scheme, the resulting trajectories might be analysed, as an ensemble or individually, to provide insight into the nonadiabatic process of interest. As a part of MLatom, it is possible to carry out such an analysis easily using several built-in functions. For plotting of one trajectory or an entire ensemble, only one line of Python code is needed, while the flexibility is maintained to focus on a certain region.

\begin{figure}
    \centering
    \begin{subfigure}{.48\textwidth}
        \centering
        \includegraphics[width=\textwidth]{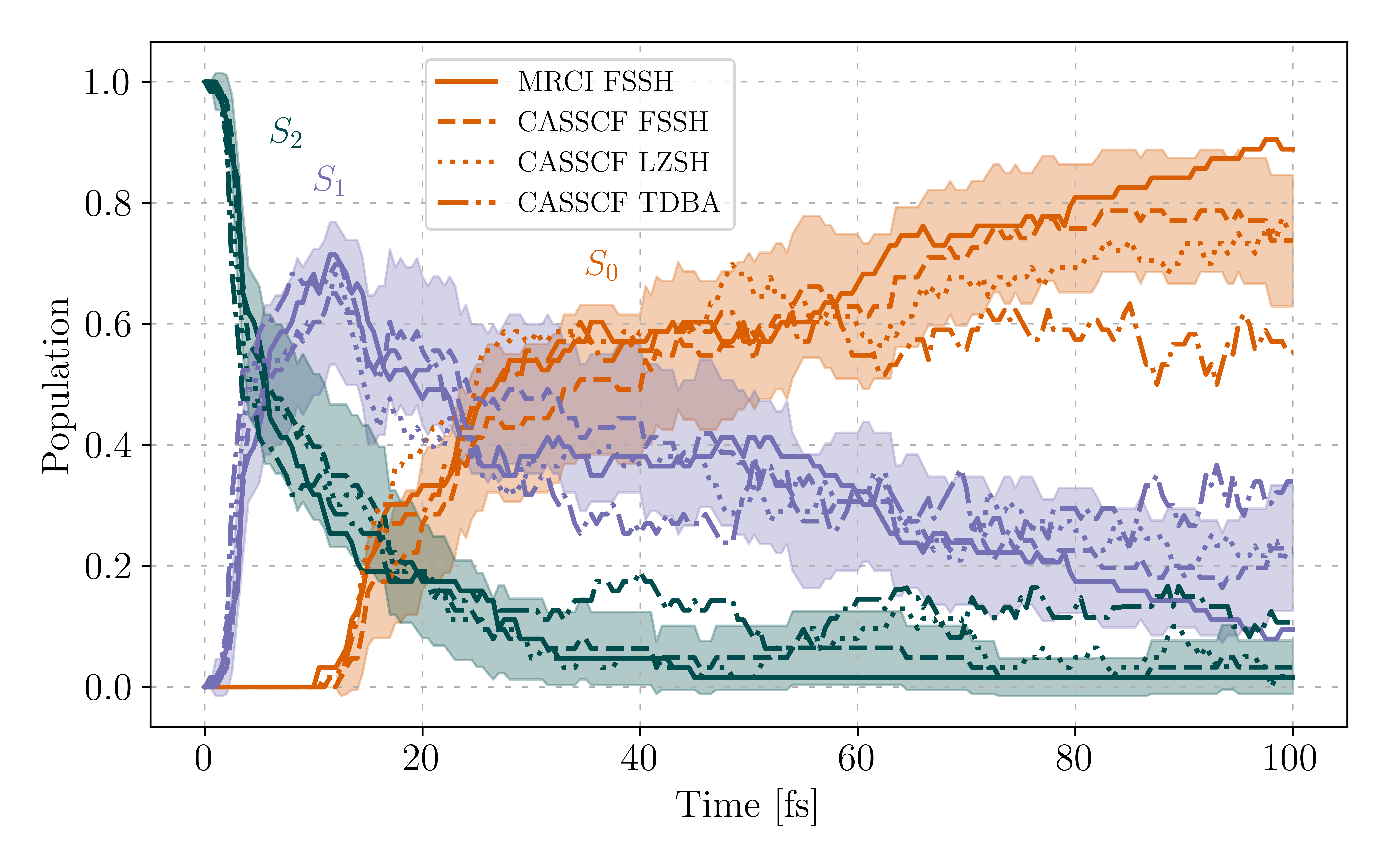}
    \end{subfigure}
    \begin{subfigure}{.48\textwidth}
        \centering
        \includegraphics[width=\textwidth]{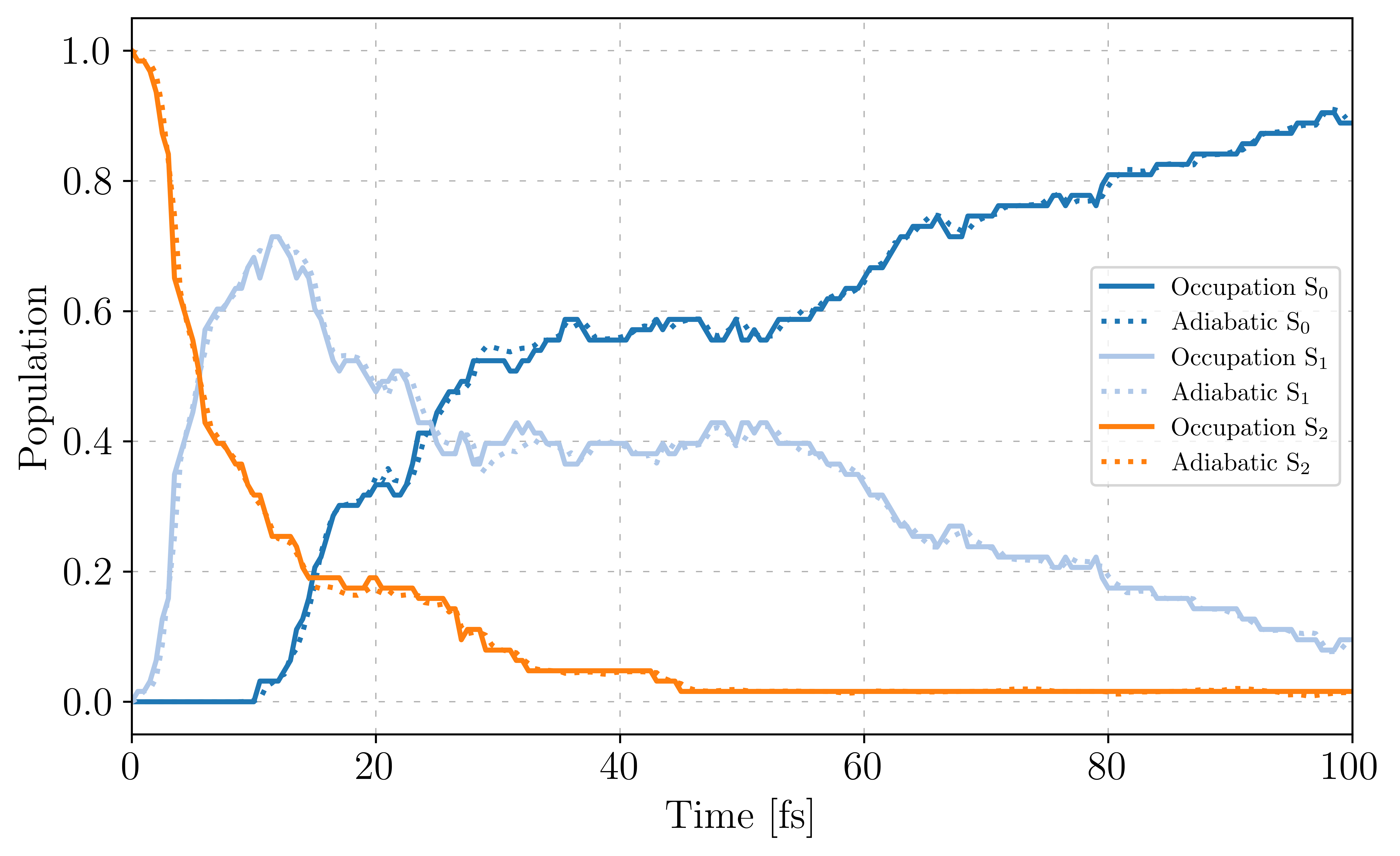}
    \end{subfigure}
    \caption{(a) Population plot (occupation) for 63 trajectories of the methylenimmonium cation computed at the SA-3-CASSCF(12,8)/6-31G* and MR-CISD(6,4)/SA-3-CASSCF(6,4)/6-31G* levels of theory. The two methods agree qualitatively. When CASSCF is employed with the LZSH scheme, the population remains within the associated confidence interval. In contrast, the TDBA exhibits a noticeable deviation after approximately 60 fs. (b) Comparison of the state occupations and the average adiabatic population obtained via the \texttt{internal\_consistency\_check()} method for FSSH trajectories at the MRCI level of theory.}
    \label{fig:analysis}
\end{figure}

We demonstrate those MLatom's abilities on the third example of the methylenimmonium cation, the smallest representative of a protonated Schiff base. After photoexcitation into $S_2$ the system relaxes into $S_1$ within just 10 fs followed by slower relaxation into $S_0$. The $S_2/S_1$ CI is associated with CN stretching and $S_1/S_0$ CI with torsion around CN bond\cite{Barbatti2006}. The MLatom is interfaced with Columbus, so the simulations can be run at both, CASSCF or MRCI levels of theory (indicated by \texttt{level\_of\_theory} keyword). We therefore compare previously used settings and the calculations were carried out at the SA-3-CASSCF(12,8)/6-31G* and MR-CISD(6,4)/SA-3-CASSCF(6,4)/6-31G* level of theory. The MRCI simulations started at $S_2$ with a time step of 0.5 fs up to 100 fs. We used only 63 trajectories out of 200 due to the dissociation or convergence problems, which is consistent with the previous studies\cite{Westermayr2019}. The CASSCF trajectories started from the same initial conditions. The resulting population is shown in Fig.~\ref{fig:analysis}; both methods agree qualitatively. 

When analysing an ensemble, the population can be automatically calculated using the \texttt{plot\_population()} method, which creates a text file containing the corresponding data. The population data can then be easily loaded and plotted together for various surface hopping schemes, QC methods or ML models as demonstrated for fulvene and ferro-wire examples. The results for MRCI and CASSCF are shown in Fig.~\ref{fig:analysis}a. Another analysis possible with FSSH or TDBA trajectories is a comparison between the occupation and average adiabatic population (as in Eq.~\ref{eq:population}). This can be done using the \texttt{internal\_consistency\_check()} method, which produces the plot shown in Fig.~\ref{fig:analysis}b.

When comparing two ensembles of trajectories calculated using the QC method and the ML model, it can be useful to plot specific degrees of freedom or analyse hopping geometries. Unfortunately, this depends on the definition of the configurational change of interest, which is highly system-specific. It is therefore possible to extract hopping structures and apply unsupervised learning techniques, such as those available in the Scikit-learn\cite{scikit-learn} library. In MLatom, trajectories can be visualized via the Python API using the \texttt{view()} method with \texttt{py3Dmol}\cite{py3Dmol}, or coordinates can be written to a text file and analysed with external tools. Additionally, MLatom provides methods to calculate bond lengths, angles, and dihedral angles, which simplifies the automatic plotting analysis.

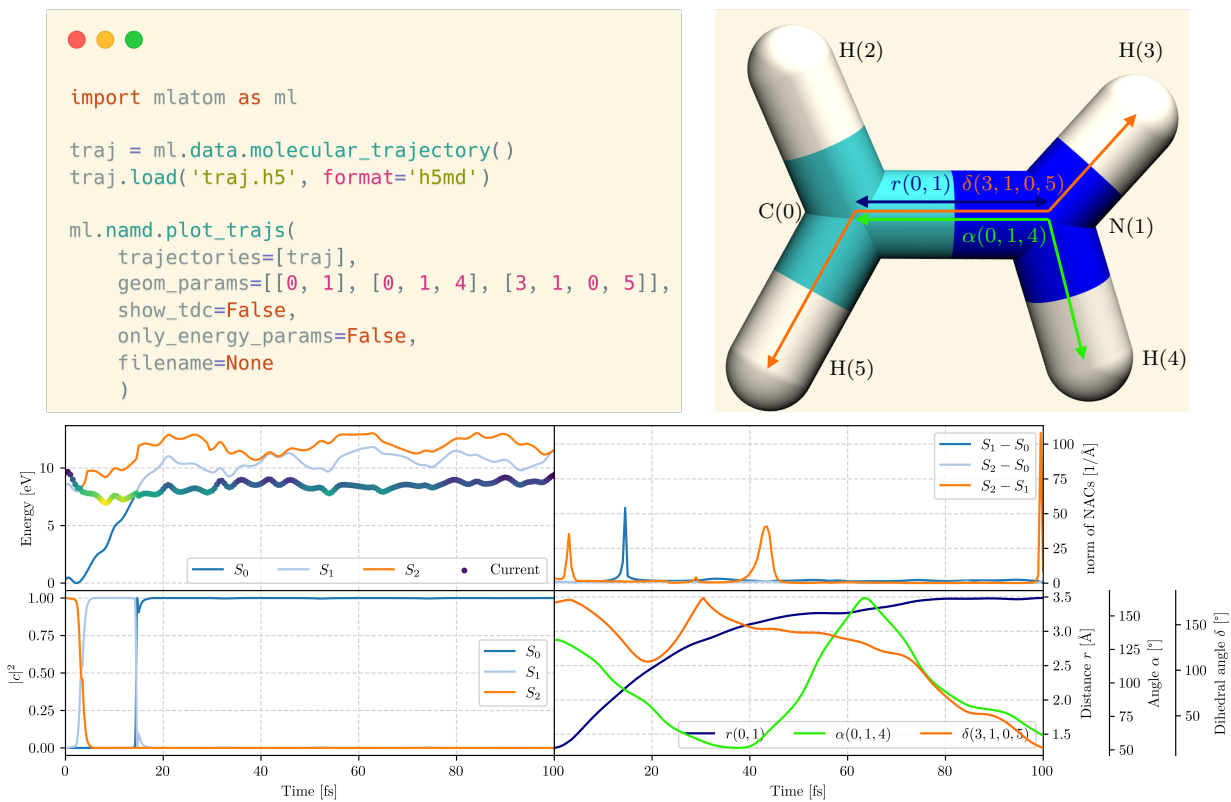
\begin{figure}
    \centering
    \includegraphics[height=13em]{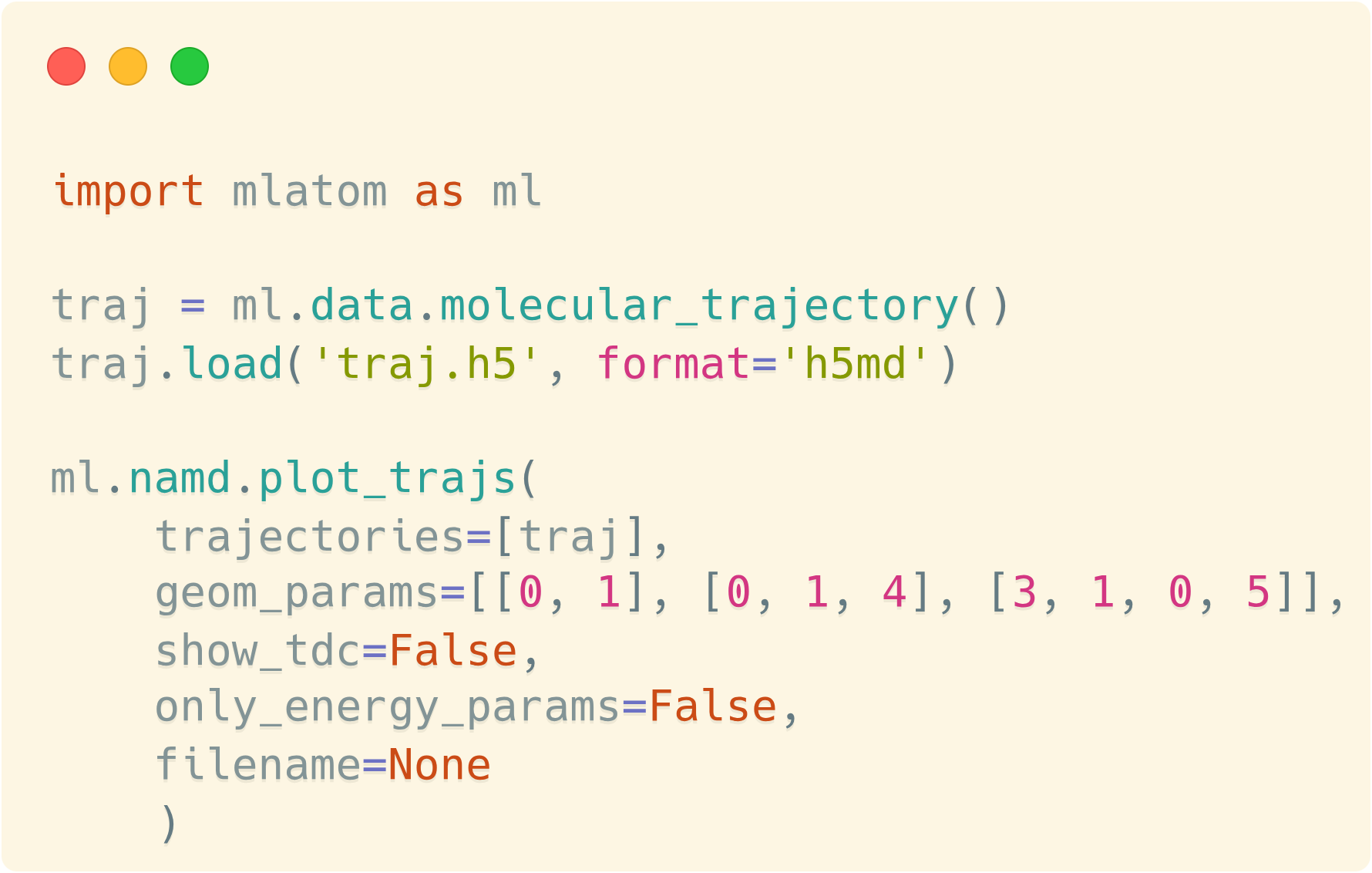}
    \hspace{0.1em}
    \begin{overpic}[height=13em]{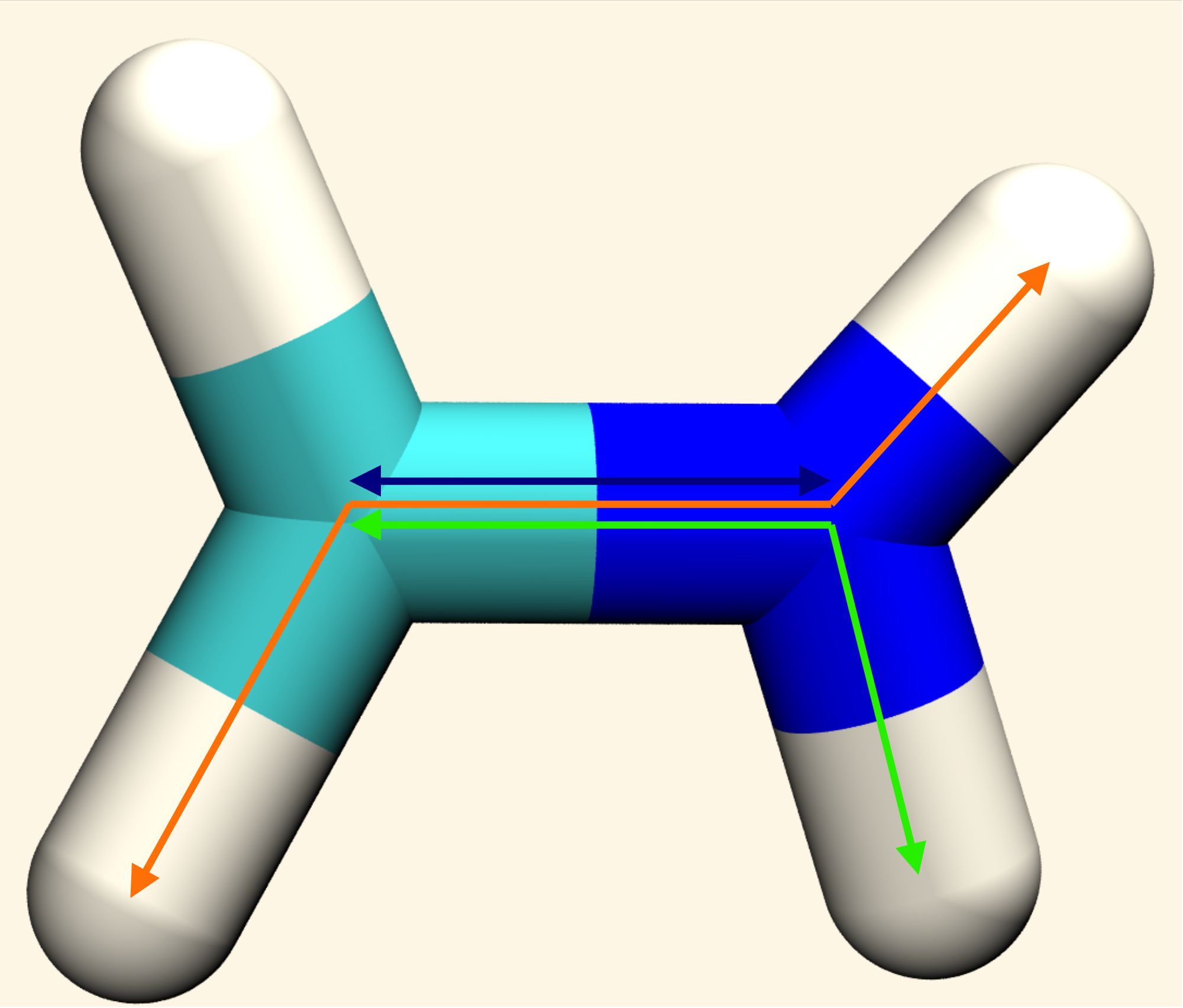}
        \put(37,47){\scriptsize $\textcolor[HTML]{010180}{r(0,1)}$}
        \put(52,36){\scriptsize $\textcolor[HTML]{26f000}{\alpha(0,1,4)}$}
        \put(52,47){\scriptsize $\textcolor[HTML]{fe7007}{\delta(3,1,0,5)}$}
        \put(9,41){\scriptsize C(0)}
        \put(83,38){\scriptsize N(1)}
        \put(90,10){\scriptsize H(4)}
        \put(85,75){\scriptsize H(3)}
        \put(26,75){\scriptsize H(2)}
        \put(24,8){\scriptsize H(5)}
    \end{overpic}
    \vspace{1em}
    \includegraphics[width=\textwidth]{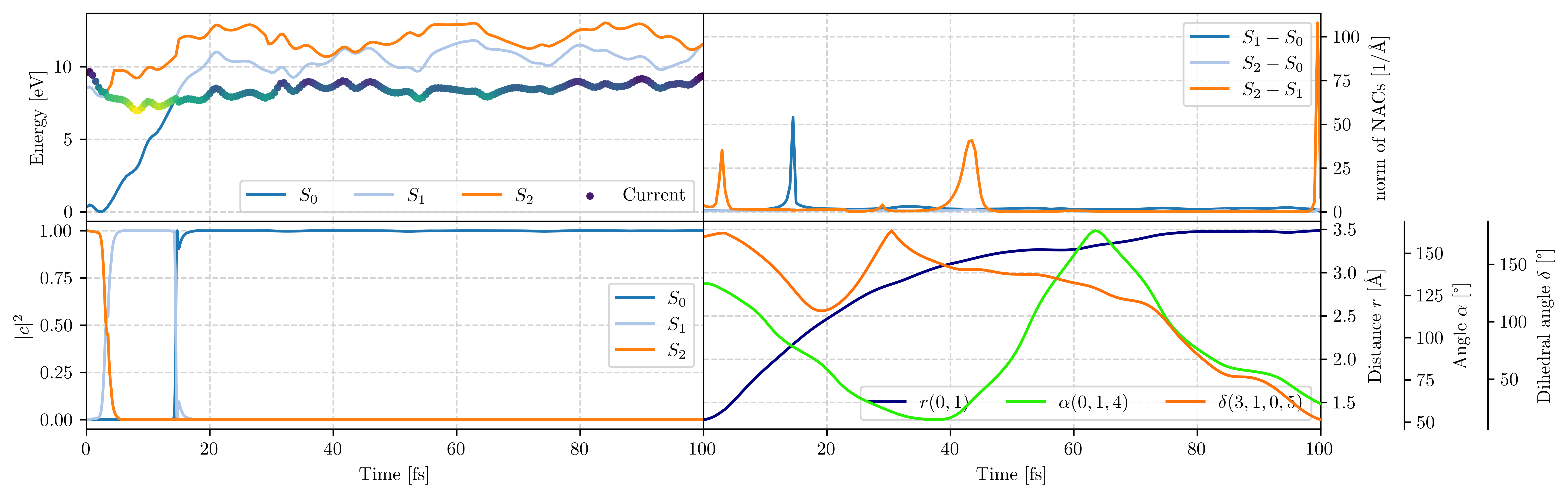}
    \caption{Top left: Code snippet showing the use of the \texttt{plot\_trajs()} method for a single FSSH trajectory of the methylenimmonium cation at the MRCI level of theory. The \texttt{geom\_params} parameter specifies selected degrees of freedom (bond length $r$, angle $\alpha$, or dihedral angle $\delta$). It is possible to plot TDCs instead of NACs using \texttt{show\_tdc} parameter, or to plot only the PESs and degrees of freedom using the \texttt{only\_energy\_params} parameter. Top right: Initial structure of the trajectory, showing the selected degrees of freedom. Bottom: Typical multiplot output of the \texttt{plot\_trajs()} method. The top-left subplot displays the time evolution of the potential energy, where markers indicate the current state and the marker colour represents the kinetic energy. The top-right subplot shows Frobenius NAC norms (when NACs are available, i.e., in the case of FSSH) or TDCs (in the case of TDBA or FSSH). The bottom-left subplot displays the evolution of adiabatic population, and the bottom-right subplot visualises the evolution of selected degrees of freedom. When NACs or TDCs are not available, as is the case of LZSH, only the top-left and bottom-right plots are shown.}
    \label{fig:traj_analysis}
\end{figure}

An analysis of a single trajectory provides insight at the atomistic level. Here we use the \texttt{plot\_trajs()} method, which loops over trajectories in a list and creates a useful figure consisting of 2 or 4 subplots. The first one (top left in Fig.~\ref{fig:traj_analysis}) contains the time evolution of potential energy, which traces the active state and kinetic energy (point colour) and from which the discontinuities in the PESs can be identified. The second one (bottom right in Fig.~\ref{fig:traj_analysis}) enables visualisation of the evolution of selected degrees of freedom (bond length, angle, or dihedral angle) specified by a list of their indices. When FSSH or TDBA surface hoppings are run, the trajectories contain data about the evolution of state coefficients (bottom left in Fig.~\ref{fig:traj_analysis}) and TDCs or NAC norms (top right in Fig.~\ref{fig:traj_analysis}). One can easily see the differences between MRCI and CASSCF trajectories, or between different surface hopping schemes. In the case of the methylenimmonium cation example, it is evident from such plots that TDBA suffers from discontinuity-induced hops, which are less common in LZSH trajectories.

This kind of analysis is especially useful when ML models are trained and tested iteratively. This provides a fast visualisation tool that speeds up the development as it requires only one line of code to produce Fig.~\ref{fig:traj_analysis}. The flexibility lies in the freedom of using corresponding methods to focus on a certain region by restricting the axis range of the plot. All the data are loaded and extracted internally (for demonstration, see the examples in the Jupyter Notebook corresponding to this work: \url{https://github.com/JakubMartinka/FSSH-in-MLatom}). The insight from our analysis tools helps to understand challenges in ML-accelerated NAMD, but also provides a standardised way of comparing different surface hopping schemes. 

\section{Conclusions and Outlook}
In this work, we present a versatile framework for conducting NAMD simulations using Tully's FSSH and its approximate variant employing TDBA couplings newly implemented in MLatom, extending the previously implemented LZSH scheme. We have demonstrated its capabilities on three representative examples of fulvene, a molecular ferro-wire and methylenimmonium cation, each reflecting the key stages in a typical NAMD study.

The first stage is the selection of a QC method or an ML model. We showcased the use of custom models that allow users to define rules imposed on the method providing energies, gradients, and NACs. This setup is useful for understanding the limitations of newly developed ML models and for benchmarking different surface hopping schemes. We showed that they can significantly accelerate simulations without loss of accuracy or, when using ML models, provide efficient insight into the behavour of surface hopping algorithms. 

The second stage is the choice of a surface hopping scheme, which depends on the availability of NACs. Since many QC methods or ML models do not supply them, users are restricted to curvature-driven schemes. We compared TDBA and LZSH against FSSH as a reference. In all cases, LZSH outperforms TDBA, noting that no special conditions or hop-blocking algorithms were applied. The TDBA scheme can even lead to erroneous hopping, which can be mitigated by applying the aforementioned conditions or by enabling hopping only in the nonadiabatic region. Although TDBA scheme is frequently used in ML-accelerated NAMD, based on evidence from the literature and our results, we strongly recommend using LZSH instead of TDBA.

The final stage of a NAMD study is the analysis. We introduced several easy-to-use plotting tools that provide chemical insight and support benchmarking and development workflows. Trajectories are stored in binary H5MD format and reloaded from disk when needed. They can be analysed either individually or as an ensemble, with full flexibility to focus on specific regions while all the data manipulation is handled automatically.

Overall, these developments broaden the scope of NAMD simulations available through MLatom. The framework offers flexible access to surface hopping schemes, including FSSH, TDBA, and LZSH. Combined with custom models and comprehensive analysis tools, MLatom lowers the practical barrier to exploring excited-state dynamics using both traditional QC and ML approaches, enabling more efficient and insightful research.

\subsection{Code availability}
The code is available in the open-source MLatom under the MIT license as described at \url{https://github.com/dralgroup/mlatom}. The instructions to reproduce this work can be found in Jupyter Notebook on the GitHub repository associated with this work at \url{https://github.com/JakubMartinka/FSSH-in-MLatom}.

\subsection{Authors contributions}
J.M. implemented the final version of FSSH, TDBA and plotting tools, performed final calculations and their analysis, and wrote the original manuscript.
M.M. obtained the training set for fulvene dynamics, trained the MS-ANI model for PESs, prepared the MNDO inputs and contributed to the analysis and debugging of the FSSH code.
B.M. contributed to the interfaces implementation and debugging of the FSSH code.
J.P. co-designed and co-conceived the project, contributed to the result analysis and interpretation, co-supervised research, and secured funding.
P.O.D. co-designed and co-conceived the project, contributed to the result analysis and interpretation, co-supervised research, and secured funding.
All authors discussed the results and revised the manuscript.

\begin{acknowledgement}
The work of the Czech team has been supported by the \textit{Czech Science Foundation} Grant \mbox{23-06364S}, the \textit{Charles University} (project GAUK 6224) and
by the Advanced Multiscale Materials for Key Enabling Technologies project of the \textit{Ministry of Education, Youth, and Sports of the Czech Republic}, project No. CZ.02.01.01/00/22\_008/0004558, co-funded by the European Union.
We also highly appreciate the generous admission to computing facilities owned by parties and projects contributing to the National Grid Infrastructure MetaCentrum provided under the program ``Projects of Large Infrastructure for Research, Development, and Innovations ''(no. LM2010005) and computer time provided by the IT4I supercomputing center (Project ID:90254) supported by the Ministry of Education, Youth and Sports.

P.O.D. acknowledges funding by the projects for International Senior Scientists (Project No.: W2531013) and for Outstanding Youth Scholars (Overseas, 2021) of the National Natural Science Foundation of China, via the Lab project of the State Key Laboratory of Physical Chemistry of Solid Surfaces, and Aitomistic, Shenzhen.
\end{acknowledgement}




\bibliography{ALL}

\end{document}